\documentclass{article}

\usepackage{arxiv}

\usepackage[utf8]{inputenc} 
\usepackage[T1]{fontenc}    
\usepackage{hyperref}       
\usepackage{url}            
\usepackage{booktabs}       
\usepackage{amsfonts}       
\usepackage{nicefrac}       
\usepackage{microtype}      
\usepackage{lipsum}

\usepackage{graphicx}
\usepackage{subfigure}

\usepackage{float}
\usepackage{amsmath,amssymb,amsthm,bm}
\usepackage{xcolor}
\usepackage{ulem}
\usepackage{multirow}
\usepackage{makecell}

\DeclareGraphicsExtensions{-eps-converted-to.pdf}
\title{Machine Learning for Vortex Induced Vibration in Turbulent Flow}
\author{
  Xiaodong Bai $^{1}$\\
Ministry-of-Education Key Laboratory\\ of Coastal Disaster and Defence, \\Hohai University, Nanjing 210098, China\\
  \texttt{xdbai@hhu.edu.cn} \\
  \And
  Wei Zhang $^{2,*}$\\
  Science and Technology on Water-Jet Propulsion Laboratory\\
  Marine design and research institute of China\\
  Shanghai 200011, China\\
  \texttt{Corresponding Author : waynezw0618@163.com} \\
}

\begin{document}
\maketitle

\begin{abstract}
Vortex induced vibration (VIV) occurs when vortex shedding frequency falls close to the natural frequency of a structure. Investigation on VIV is of great value in disaster mitigation, energy extraction and other applications. Following recent development in machine learning on VIV in laminar flow, this study extends it to the turbulent region by employing the state-of-the-art parameterised Navier-Stokes equations based physics informed neural network (PNS-PINN). Turbulent flow with Reynolds number $Re$ = $10^4$, passing through a cylinder undergoing VIV motion, was considered as an example. Within the PNS-PINN, an artificial viscosity $\nu_t$ is introduced in the Navier-Stokes equations and treated as a hidden output variable. A recently developed Navier-Stokes equations based PINN, termed NSFnets (Jin \textit{et al.}, J COMPUT PHYS 426: 109951, 2021), was also considered for comparison. Training datasets of scattered velocity and dye trace concentration snapshots, from computational fluid dynamics (CFD) simulations, were prepared for both the PNS-PINN and NSFnets. Results show that, compared with the NSFnets, the PNS-PINN is more effective in inferring and reconstructing turbulent flow under VIV circumstances. The PNS-PINN also shows the capability to deal with unsteady and multi-scale flows in VIV. Inspired by the Reynolds-Averaged Navier-Stokes (RANS) formulation, by implanting an additional artificial viscosity, the PNS-PINN is free of complex turbulence model closure, thus may be applied for more general and complex turbulent flows.
\end{abstract}


\section{Introduction}

Flow induced vibration is widespread in nature, industry and everyday human life. VIV, as a branch of the flow induced vibration, occurs when the natural vibration frequency of a structure is close to the vortex shedding frequency. According to the conditions of the reduced velocity, the structural dynamic characteristics, the Reynolds number and other factors, VIV may cause large amplitude vibration of structures \cite{Blevins1990, Williamson2004}. Research on VIV has great significance for disaster mitigation in ocean and civil engineering practices \cite{Laima2013, Liu2020, Wang2018}, and even the sustainable energy extraction \cite{Bernitsas2008, Wang2020b}. CFD, as an efficient way to study VIV, is mostly limited to flows with low Reynolds number \cite{Evangelinos1999, Prasanth2008, Griffith2017, Sabino2020}, due to the high computational cost introduced by mesh refinement and update of the fluid-structure interaction (FSI), and more importantly the stiffness issue induced by high Reynolds number $Re$. For VIV in turbulent flow with high $Re$, experimental measurement exists an important approach for investigation. With experiment, motions of the structure can be measured precisely. Meanwhile, scattered velocity or dye trace distribution around the VIV structure in a small region can be obtained experimentally, for example, via the particle image velocimetry \cite{Raffel1998} or the Schlieren image velocimetry measurements \cite{Jonassen2006, Hargather2011}. Yet it is also acknowledged that the pressure distribution is difficult to acquire \cite{VanGent2017, ZhangJC2020}. Thus, how to accurately reconstruct the pressure field and thereafter predict the forces according to limited available data in finite time-space coverage is of interest and importance. 

With the proliferation of large-scale computing capability witnessed in recent years, the machine learning technique provides a powerful alternative to explore complex fluid mechanics. Modeling of turbulent flow based on data-driven machine learning, for example by developing or improving turbulence closure models, has gained considerable progress. Ling \textit{et al.} \cite{Ling2016} devised a deep neural network incorporating the Galilean invariance, which provides better representation for the Reynolds stress compared with that from baseline RANS models. Xiao \textit{et al.} \cite{Xiao2016} developed a data-driven and physics-informed Bayesian framework for quantifying and reducing uncertainties in RANS simulations, and show that the proposed framework delivered better quantities even with sparse observations. To improve the accuracy of RANS models, Wang \textit{et al.} \cite{Wang2017} established a machine learning based on the random forest method, to train the discrepancy function with data from direct numerical simulation (DNS). Weatheritt and Sandberg \cite{Weatheritt2017} developed a multi-dimensional gene expression programming (MGEP) algorithm and introduced an additional algebra term to reduce discrepancy and to calibrate close coefficients of quadratic eddy-viscosity models. Based on MGEP and DNS data, Zhang \textit{et al.} \cite{Zhang2019} considered a constraint-free model for Reynolds stress within a rotating frame of reference, and studied the anisotropic behavior of Reynolds stress. Zhu \textit{et al.} \cite{Zhu2019} employed a one-layer radial basis function neural network to predict flows around airfoils, and demonstrated higher efficiency of the network compared to the original Spallart-Allmaras model. Machine learning was also employed early in works of Sarghini \textit{et al.} \cite{Sarghini2003} and Barone \textit{et al.} \cite{Barone2017} to determine the large-eddy simulations (LES) closures and to reduce computational cost. Schoepplein \textit{et al.}  \cite{Schoepplein2018} extended the MGEP in RANS to model subgrid stresses, and demonstrated that evolutionary algorithms can successfully be incorporated in the framework of LES. Zhou  \textit{et al.} \cite{Zhou2019} established an artificial neural network to investigate the relation between the resolved flow field and the subgrid-scale stress tensor, and found that the correlation of the new subgrid-scale  stress tensor and the DNS data was mostly over 0.9 and the energy transfer rate can be well predicted at various Reynolds numbers. 

The above mentioned investigations are mostly application-oriented, and focused on specific turbulence models. Machine learning can be more intelligent if the neural network is well informed by proper algorithms. Recently, Raissi \textit{et al.} \cite{Raissi2019a} introduced physics informed neural network (PINN), which was designed to embed underlying physics laws that govern the known and unknown data, thus solutions of the forward or inverse problems can be inferred with higher fidelity. The PINN was later extended by the authors to explore more complex fluid problems, such as hemodynamics \cite{Raissi2020}. With that, Jin \textit{et al.} \cite{Jin2020} proposed the NSFnets, which incorporate the Navier-Stokes equations based on the PINN, to simulate laminar and turbulent channel flows with DNS. Bai \textit{et al.} \cite{Bai2020} applied the PINN with various attempts on flow data assimilation, concerning flow inference, reconstruction, identification for both laminar and turbulent flow scenarios. Therein, both the discrete Boltzmann equation and the Navier-Stokes equations were employed as physics constraints. To explore missing fluid dynamics with high Reynolds numbers, a parameterised Navier-Stokes equations based PINN (PNS-PINN) was proposed recently by Xu \textit{et al.} \cite{Xu2020}. Inspired by the RANS equation, an artificial viscosity $\nu_t$ was introduced and trained as a hidden variable in the system. The parameterised system is free of extra turbulence modeling or additional closure calibrations for data assimilation. 

Machine learning has also been used for VIV studies with low Reynolds number laminar flow, to infer flow field and predict the forces acting on structure. In Raissi  \textit{et al.}'s recent work \cite{Raissi2019}, VIV of a freely vibrating cylinder in laminar flow with $Re$ = 100 was studied via the deep neural network. Therein, the incompressible Navier–Stokes equations considering the dynamic motion  of the VIV cylinder were integrated in the loss function. Scattered velocity data of flow close to the cylinder were considered as the training dataset. Their results show that the PINN can be employed to predict the drag and lift forces with promising precision. To the best of our knowledge, the possibility of exploring VIV in turbulent flow via machine learning has not been demonstrated yet, probably due to high Reynolds number stiffness. 

Inspired by the works mentioned above, especially the PINN configuration \cite{Raissi2019} and the parameterised governing systems \cite{Xu2020}, this paper investigates the VIV in turbulent flow via the PNS-PINN. In Section 2, the parameterised Navier-Stokes equations are introduced, followed by details of the VIV system, the configuration and dataset preparation of the PNS-PINN, as well as a normal Navier-Stokes equation based PINN (NSFnets). Flow inference, pressure field reconstruction and force prediction of the VIV using the PINNs with various datasets are reported in Section 3. The performances of the PINNs based on two datasets (velocity and dye trace) are also analyzed and evaluated. Concluding remarks are made in Section 4.

\section{Methods}
\subsection{Parameterised Navier-Stokes equations}

Incompressible flows are governed by the Navier-Stokes equations as follows,
\begin{equation}
\begin{aligned}
&\frac{\partial \bf{u}}{\partial t}-\nu \nabla^2 \bf{u} + \bf{u} \cdot \nabla \bf{u} +\nabla p =\bf{f}\\
&\nabla \cdot \bf u= \text{0}
\end{aligned}
  \label{Eq:1}
\end{equation}
where, $\bf{u}$ = $ [u,  v,  w]^\text{T}$, $p$ and $\nu$ are the flow velocity, pressure and kinematic viscosity respectively; $\bf{f}$ is the force acting on a finite volume.

Ideally, incompressible flows that are governed by Eq.~(\ref{Eq:1}) can be resolved numerically with properly defined initial and boundary conditions. However, when $\nu$ is of very small value, as the case for most turbulent flows, direct solution for the above partial differential equations becomes extremely difficult due to not only large computational cost, but more importantly, the illness or stiffness of the algebraic matrices involved. For industrial practices, despite of well-known limitations, RANS models with various formulas still serve with great reputation when dealing with complex turbulent flow scenarios  \cite{Wilcox2006}.  To eliminate the temporal dependency, the RANS employs an ensemble-averaging process, which leads to unclosed Reynolds stress. The key point of the enclosure and modeling of the Reynolds stress is to treat it as a function of the gradient of average flow velocities. Inspired by that, an artificial viscosity is introduced in the governing equations for PINN \cite{Xu2020}, and the parameterised system reads as following,
\begin{equation}
\begin{aligned}
&\frac{\partial \bf{\overline{u}}}{\partial t}-(\nu + \nu_t)\nabla^2 \bf{\overline{u}} + \bf{\overline{u}} \cdot \nabla \bf{\overline{u}} +\nabla \overline{p} =\bf{\overline{f}}\\
&\nabla \cdot \bf \overline{u}= \text{0},
\end{aligned}
  \label{Eq:2}
\end{equation}

In the above equation, the overline label $\overline{(\cdot)}$ denotes an operator that can be either of ensemble average processing in RANS or filtering processing in LES (large eddy simulation), so that  the PNS-PINN can be used to treat multi-scale flow data of turbulent flow.; $\nu_t$ is the introduced artificial viscosity and  plays as the key parameter. 

\subsection{Details of the VIV}
The VIV motion of a circular cylinder and the surrounding turbulent flow are considered in this study. The structure motion is modeled as an elastically supported mass-spring-damper system, with only the cross-flow motion allowed. The governing equation of the cylinder motion $\eta$ is thus given by Eq.~(\ref{Eq:3}), same as that used in Ref. \cite{Raissi2019}:
\begin{equation}
    m \frac{\text{d}^2 \eta}{\text{d} t^2} + b \frac{\text{d} \eta}{\text{d} t} +k \eta =F_L(t)
    \label{Eq:3}
\end{equation}
Parameters $m$, $b$ and $k$ are the mass, damping and stiffness of the elastically supported system  respectively. $F_L$ is the lift force as a result of the FSI.

CFD simulations of the VIV were carried out, and the numerical results were used as the training datasets. It should be noted that dataset from experimental measurement can also be used for the training. Two-dimensional fluid motions were simulated via the solver, pimpleDyMFoam, implemented in OpenFOAM$^\circledR$, which is an open source framework of finite volume method. Two turbulence models, including the SST (shear stress transport) $k-\omega$  model and the SST-SAS (scale adaptive simulation) mixed model were employed. The former model is robust for external aerodynamics, while the latter one is an improved version at dealing with multi-scale wake flows, as discussed in Refs. \cite{Menter1994} and \cite{Menter2003}. 

The computational domain is a circular area with a diameter of 70$D$. $D$ is the diameter of the cylinder, whose centroid point is located at $O$ (0, 0). The inlet flow is applied on the left half part of the computational boundary with a Dirichlet boundary condition $\bm{u}=$ ($U_\infty, 0)$, while the right half is defined as an outlet boundary with zero-gradient pressure. The cylinder boundary is non-slip. The Reynolds number, $Re$ = $U_\infty D/ \nu$, is $10^4$. The cylinder wall is prescribed with a constant concentration value $c_0$ = 10, which will be transported downstream due to both the advection and diffusion. The governing P\'{e}clet number $Pe$ = $U_\infty D/\nu_\text{diff}$ = 90 is adopted, where $\nu_\text{diff}$ is the diffusivity of the concentration. The parameters for the cylinder are $m$ = 2, $b$ = 0.084 and $k$ = 2.202, same as those used in Ref. \cite{Raissi2019}, and the natural frequency of the cylinder vibration is $f_0$ = 0.167 Hz. The reduced velocity can be determined as $U_r$ = $U_\infty / D f_0 =6.0$, with $D$ and $U_\infty$ the characteristic length and velocity respectively. A dynamic mesh scheme is adopted to update the computational mesh as the cylinder boundary moves due to the VIV. 

\begin{figure}[htbp]
  \centering
    \includegraphics[trim= 25mm 25mm 18mm 15mm,clip, width=1.0\textwidth]{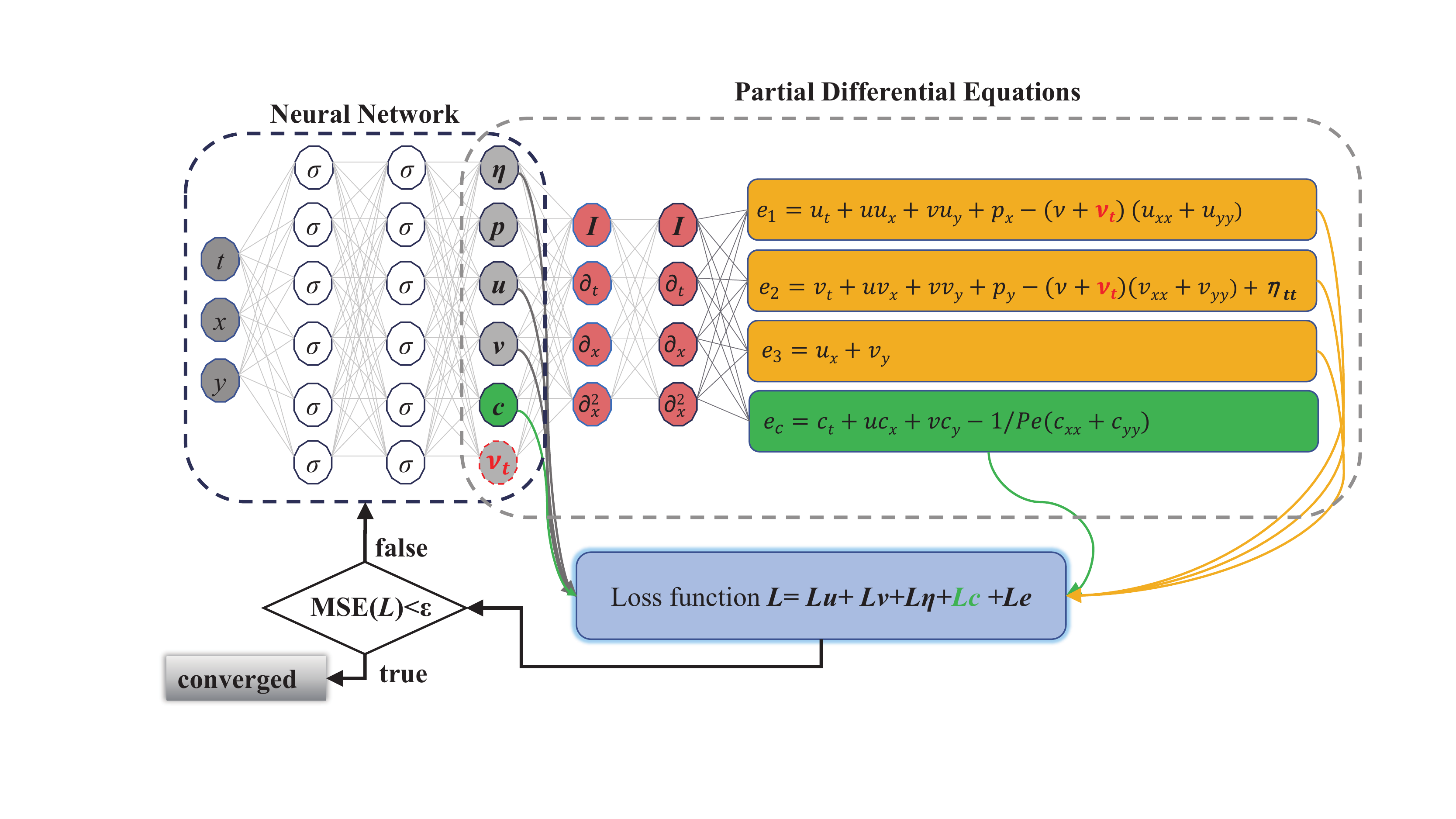}
  \caption{Configurations of the PNS-PINN and the NSFnets.}
  \label{Fig:1}
\end{figure}

\subsection{PINNs and the training dataset}
The purpose of this study is to infer the turbulent flow and the forces acting on the moving structure, based on scattered  \{${t^n, \eta^n}$\}, and scattered flow information such as velocity \{$t^n, x^n, y^n, u^n, v^n$\} or dye trace \{$t^n, x^n, y^n, c^n$\} in the near vicinity of the VIV cylinder. Here, $t$ is the time; $x$ and $y$ are the spatial coordinates; $u$ and $v$ are the velocity components; $n=1:N$, 
where $N$ is total number of data points used for training. The PINN setup,  is shown in Fig.~\ref{Fig:1}. For VIV applications, we incorporated the cylinder motion into the momentum equation shown in Eq.~(\ref{Eq:2}). Depending on whether the artificial viscosity $\nu_t$ is involved or not, the PINN can be PNS-PINN (with red labels in Fig.~\ref{Fig:1}) or NSFnets. In PNS-PINN,the artificial viscosity $\nu_t$ is treated as an output variable. For the cases based on dye trace, the concentration of the dye trace \{$t^n, x^n, y^n, c^n$\} and the cylinder motion \{${t^n, \eta^n}$\} are used as training dataset; while for cases based on velocity, the dataset contains the velocity \{$t^n, x^n, y^n, u^n, v^n$\} and also the cylinder motion. The PINN variant for the latter can be reduced by eliminating the corresponding parts involving the concentration $c$  in Fig.~\ref{Fig:1} .

Particularly, 12 hidden layers, with 32 neuros for each layer are considered. The activation function $\sigma$ is of sinusoidal type, while the temporal and spatial differential operations $\partial / \partial t$, $\partial / \partial x$, $\partial^2 / \partial x^2$ are processed with automatic differentiation formulated in TensorFlow \cite{Abadi2016}. Firstly, the loss function induced by physics governing equations of the above PINNs  consists of four parts. Therein, $e_1$, $e_2$ and $e_3$ are contributed by the streamwise velocity component $u$, the cross-flow velocity component $v$, and the mass conservation respectively. $e_c$ is for the transport of the concentration scalar if dye trace dataset is used for training. To be more specific, they read as following, 
\begin{equation}
\begin{aligned}
e_1 &=\frac{\partial u}{\partial t}+u \frac{\partial u}{\partial x}+v \frac{\partial u}{\partial y}+\frac{\partial p}{\partial x}-(\nu+\textcolor{red}{\nu_t})( \frac{\partial^2 u}{\partial x^2}+\frac{\partial^2 u}{\partial y^2}),\\
e_2 &=\frac{\partial v}{\partial t}+u \frac{\partial v}{\partial x}+v \frac{\partial v}{\partial y}+\frac{\partial p}{\partial y}-(\nu+\textcolor{red}{\nu_t})( \frac{\partial^2 v}{\partial x^2}+\frac{\partial^2 v}{\partial y^2})+ \frac{\partial^2 \eta}{\partial t^2},\\
e_3 &= \frac{\partial u}{\partial x}+\frac{\partial v}{\partial y},\\
e_c &=\frac{\partial c}{\partial t}+u \frac{\partial c}{\partial x}+v \frac{\partial c}{\partial y}-\frac{1}{Pe}(\frac{\partial^2 c}{\partial x^2}+\frac{\partial^2 c}{\partial y^2}).
\end{aligned}
  \label{Eq:4}
\end{equation}
For the simplicity of presentation, the overline symbol for the operator in Eq.\ref{Eq:2} is omitted. It should be noted that the cylinder motion $\eta$ has been incorporated in $e_2$, so that the fluid motion coordinate system is attached to the cylinder.

Then, the overall loss function is the summation of the loss of target training variables $[u, v, c, p, \eta]^\text{T}$, and the loss from the governing equations $e_i$ ($i=1,2,3$) and $e_c$. The loss function can be formulated depending on whether the velocity dataset or the dye trace dataset is provided for the machine learning, which reads as,
\begin{equation}
L = 
\begin{cases}
\begin{aligned}
	L_u+L_v+L_\eta = &\sum\limits_{n=1}^{N} \left| u(t^n, x^n, y^n)- u^n \right|^2+\sum\limits_{n=1}^{N} \left| v(t^n, x^n, y^n)- v^n \right|^2 \\ &+\sum\limits_{n=1}^{N} \left| \eta(t^n)- \eta^n \right|^2+\sum\limits_{i=1}^{3} \sum\limits_{n=1}^{N} \left| e_i \right| ^2, &\text{using velocity dataset},\\
	L_c+L_\eta = &\sum\limits_{n=1}^{N} \left| c(t^n, x^n, y^n)- c^n \right|^2+\sum\limits_{n=1}^{N} \left| \eta(t^n)- \eta^n \right|^2\\ &+\sum\limits_{i=1}^{3} \sum\limits_{n=1}^{N} \left| e_i \right| ^2+ \sum\limits_{n=1}^{N} \left| e_c \right| ^2, & \text{using dye trace dataset},
\end{aligned}	
\end{cases}
  \label{Eq:5}
\end{equation}
where, $u(t^n, x^n, y^n)$, $v(t^n, x^n, y^n)$, $c(t^n, x^n, y^n)$, $\eta(t^n)$ are the neural network training results at scattered time and spatial points. The training results for $[u, v, c, p, \eta]^\text{T}$ is determined when the loss function $L$ gets converged and minimized. 

In this study, an adaptive optimization algorithm, Adam optimizer \cite{Kingma2015}, is used to back-propagate the loss function. Due to the multi-target variables to be trained, the convergence of the loss function needs additional treatment. A modified fully-connected learning rate annealing algorithm developed by Wang \textit{et al.} \cite{Wang2020} is employed to overcome the stiffness for high Reynolds number flow. In this algorithm, two transformer networks are designed to project the inputs to a high-dimensional feature space, also a point-wise multiplication is chosen to update the hidden layers. More details can be found in Ref. \cite{Wang2020}. It is noted that the pressure distribution will be determined as a hidden variable through the physics constraint shown in Eq.~(\ref{Eq:4}) and loss function Eq.~(\ref{Eq:5})  by the neural network.

\begin{figure}[htbp]
  \centering
    \includegraphics[trim= 44.1mm 60mm 42.2mm 77mm,clip, width=1.0\textwidth]{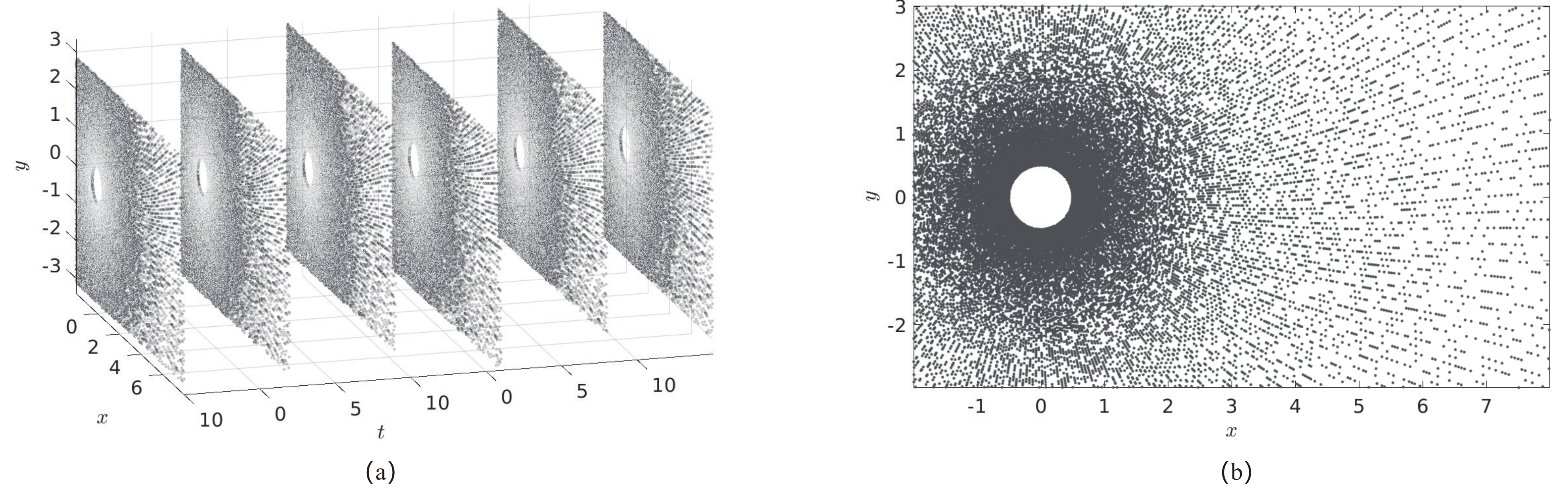}
  \caption{Temporal and spatial training data. (a) Data points at selected time instants; (b) Spatial distribution of the data points at $t$ = $\Delta t$/2.}
  \label{Fig:2}
\end{figure}

Snapshots of velocity field or the concentration distribution close to the cylinder, together with the cylinder motion are extracted from the CFD simulation results after the VIV motion of the cylinder becomes statistically stable. These snapshots are later used as the training datasets. Ultimately, the time interval of the dataset for the PINN training is determined to be $\Delta t = 14D/U_\infty$, and the time step of the training snapshots is $dt = D/50U_\infty$. Thus there are $N_s$ = 701 temporal snapshots. In the spatial domain, 4$\times 10^4$ data points are selected within a rectangular zone with $x \sim $ [-2, 8] and $y \sim $ [-3, 3] around the moving cylinder. Those scattered points have a locally dense distribution near the cylinder wall and get sparser in relatively far field. Fig.~\ref{Fig:2} illustrates the point distribution at selected time instants and a more straightforward view at time $t$ = $\Delta t$/2 in the $x-y$ plane.

A brief summary on the PINN setup and the dataset preparation can be found in Table.~\ref{Tab:1}. The training datasets and the output variables for different PINNs are listed for clarity. Depending on whether the velocity or dye trace data is used for training, the performance of the PINNs dealing with VIV in turbulent flow will be investigated and evaluated in next section.

\begin{table}[!htb]
  \centering
  \caption{Cases setup of the machine learning studies}
    \begin{tabular}{c|cc|cc|l}
\hline
    \multirow{2}[4]{*}{Dataset} & \multicolumn{2}{c}{PNS-PINN} & \multicolumn{2}{c|}{NSFnets} & \multicolumn{1}{c}{\multirow{2}[4]{*}{dataset source}} \\
\cmidrule{2-5}    \multicolumn{1}{c|}{} & training dataset & output & training dataset & output &  \\
\hline
    velocity & $[u, v, \eta]^T$ & $[u, v, p, \nu_t]^T$ & $[u, v, \eta]^T$ & $[u, v, p]^T$ & \multicolumn{1}{l}{\multirow{2}[2]{*}{ \makecell[c]{CFD simulations with SST  \\ $k-\omega$ or SST-SAS  model} }} \\
     dye trace & $[c, \eta]^T$ & $[u, v, p, c, \nu_t]^T$ & $[c, \eta]^T$ & $[u, v, p, c]^T$ &  \\
\hline
    \end{tabular}%
  \label{Tab:1}%
\end{table}%

\section{Results and Discussion}
\subsection{Machine learning using velocity dataset}
\begin{figure}[!htb]
  \centering
    \includegraphics[trim= 30mm 0mm 30mm 0mm,clip, width=0.8\textwidth]{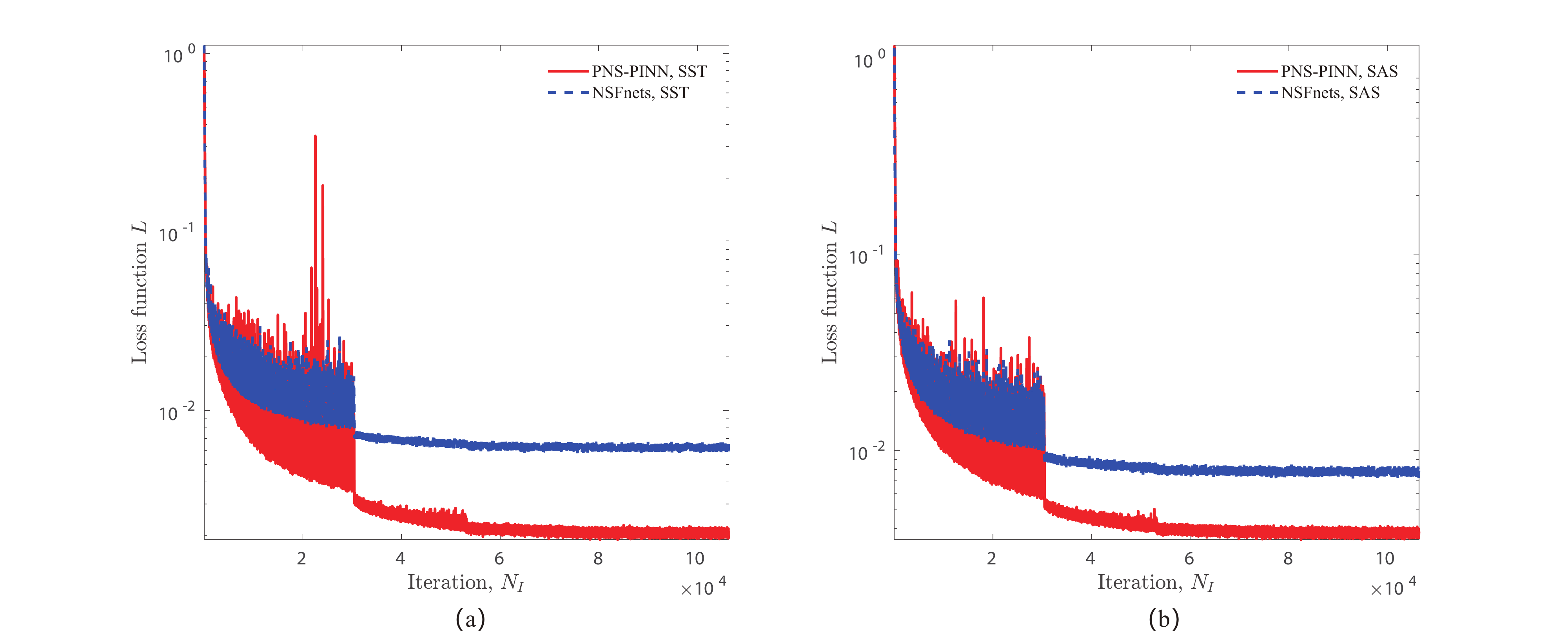}
   \caption{Time histories of the loss functions of (a) the PNS-PINN and (b) the NSFnets trainings using velocity dataset.}
  \label{Fig:3}
  \end{figure}

In this section, we analyze the machine learning results based on the scattered velocity  snapshots provided by CFD simulations. We first focus on the convergence performance of the loss function, whose time histories are given in Fig.~\ref{Fig:3}. For the training using the dataset with SST $k-\omega$ model, when the learning rate for first 3$\times 10^4$ iterations is $1.0 \times 10^{-3}$, the loss function witnesses high loss level and also large fluctuation. As the learning rate is reduced to $1.0 \times 10^{-4}$ and even lower values, the loss function declines quickly and eventually converges when the iteration number is over $10^5$. The converged loss levels for the PNS-PINN and the NSFnets using the dataset with SST $k-\omega$ model are 0.21\% and 0.62\% respectively, with the latter higher than the former and indicating probable better inferred results of the quantities involved in the loss function. Comparable but slightly higher loss levels for the trainings using the dataset with SST-SAS  model are 0.38\% and 0.78\%, which may be explained by the fact that flow simulated by the SST-SAS model are more complicated in aspect of unsteadiness.
\begin{figure}[!htb]
  \centering
    \includegraphics[trim= 40mm 5mm 5mm 0mm,clip, width=1.0\textwidth]{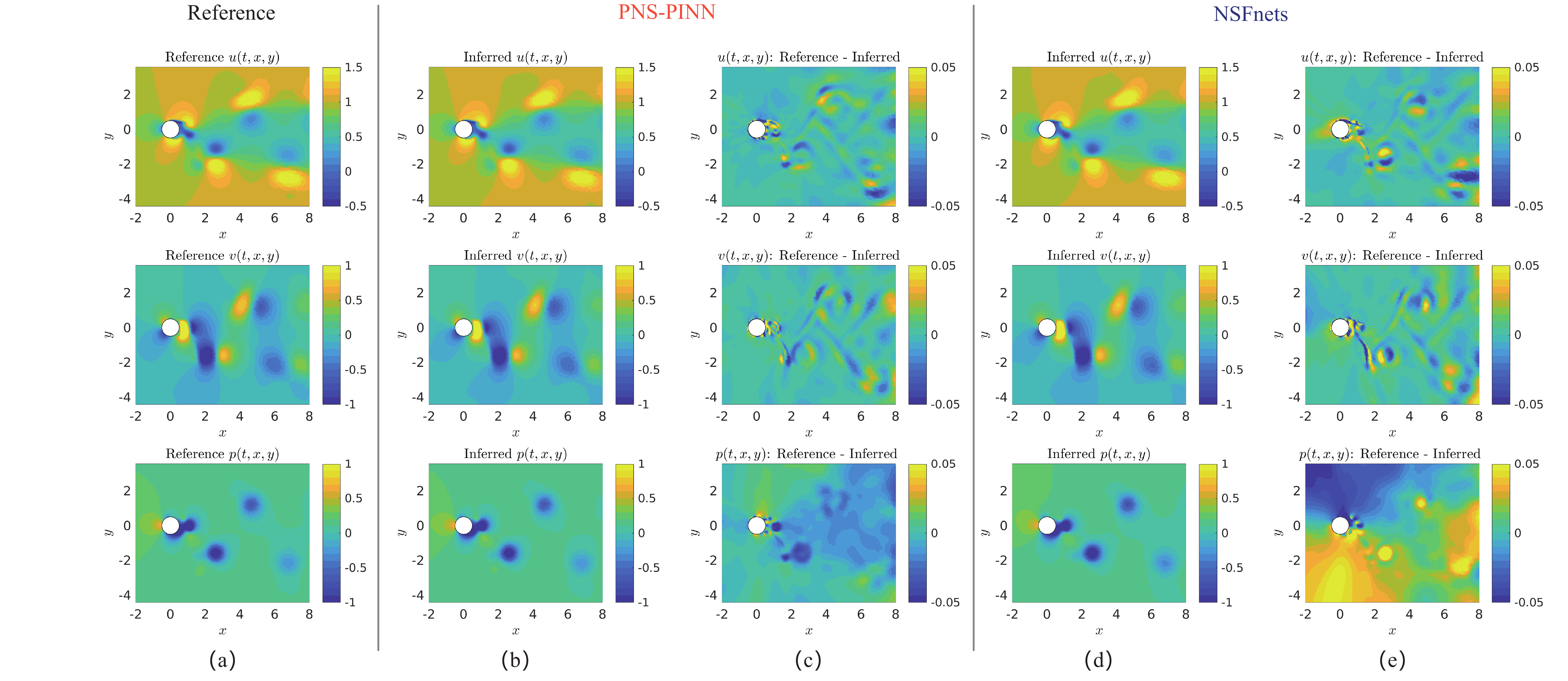}
   \caption{Flow fields inferred via the PINN and the velocity dataset with SST $k-\omega$ model: (a) reference data by CFD simulation; (b-c) inferred results via the PNS-PINN and the error distribution; (d-e) inferred results via the NSFnets and the error distribution.}
  \label{Fig:4}
  \end{figure} 
  
Based on the dataset with SST $k-\omega$ model, a series of flow data snapshots will be generated as a result of the PINN training with a total number equal to that of the dataset, \textit{i.e.}, $N_s$ = 701. Fig.~\ref{Fig:4} illustrates the inferred flow passing through the VIV cylinder at a representative time instant $t$ = $\Delta t$/2, i.e., exactly the middle time instant of the dataset coverage. The flow information by CFD simulation, shown in the first column, is labeled as the reference to evaluate the PINN training.

In Fig.~\ref{Fig:4}(b), from top to bottom, the inferred velocity ($u, v$) and pressure $p$ distributions via the PNS-PINN are presented respectively. It is found that the PNS-PINN training can reproduce the flow distributions accurately compared with the reference ones in Fig.~\ref{Eq:4}(a). Specially, the pressure field is reconstructed with satisfying approximation both in close vicinity of the cylinder and in relatively far field. Similarly, the inferred flow via the NSFnets is shown in Fig.~\ref{Fig:4}(d) for comparison. It shows that the NSFnets provides close results. The differences between the reference flow data and the inferred ones are shown in Fig.~\ref{Fig:4}(c) and (e). For the PNS-PINN, the error distributions for velocity components are in similar pattern with those of the NSFnets, except the magnitudes of the latter are augmented at the head parts of the shedding vortices and also the wall-bounded areas of the cylinder. Meanwhile, for the pressure field, the error via the PNS-PINN is obvious in the regions with vortex head and near cylinder wall. For the NSFnets, the error of the  pressure is at a  higher level in the entire domain,  especially in the vicinity of the cylinder wall. As a result, it indicates that, the pressure data can be inferred with preferable accordance via PNS-PINN, considering that the error is reduced compared with NSFnets. The successful reconstruction of the pressure field based on limited and scattered data is quite promising, considering the difficulty to acquire its distribution either in measurement (\textit{e.g.}, PIV) or in nature. 

To quantitatively evaluate the flow inference by the PINN training, a normal error coefficient $E_{L2}$ is defined as,
\begin{equation}
\begin{aligned}
 E_{L2, X}=\left[ \sum\limits_{n=1}^{N} \frac{\left|X(t^n, x^n, y^n)- X^n \right|^2}{\left|X^n \right|^2} \right] ^{1/2},
\end{aligned}
  \label{Eq:6}
\end{equation}
where, $X$ is a certain quantity of interest for analysis, \textit{i.e.}, $u$, $v$ or $p$;  $X(t^n, x^n, y^n)$ denotes the inferred data at scattered time and space ($t^n, x^n, y^n$); $X^n$ is the corresponding snapshot data of the training dataset.

\begin{figure}[!htb]
  \centering
    \includegraphics[trim= 30mm 0mm 35mm 0mm,clip, width=0.8\textwidth]{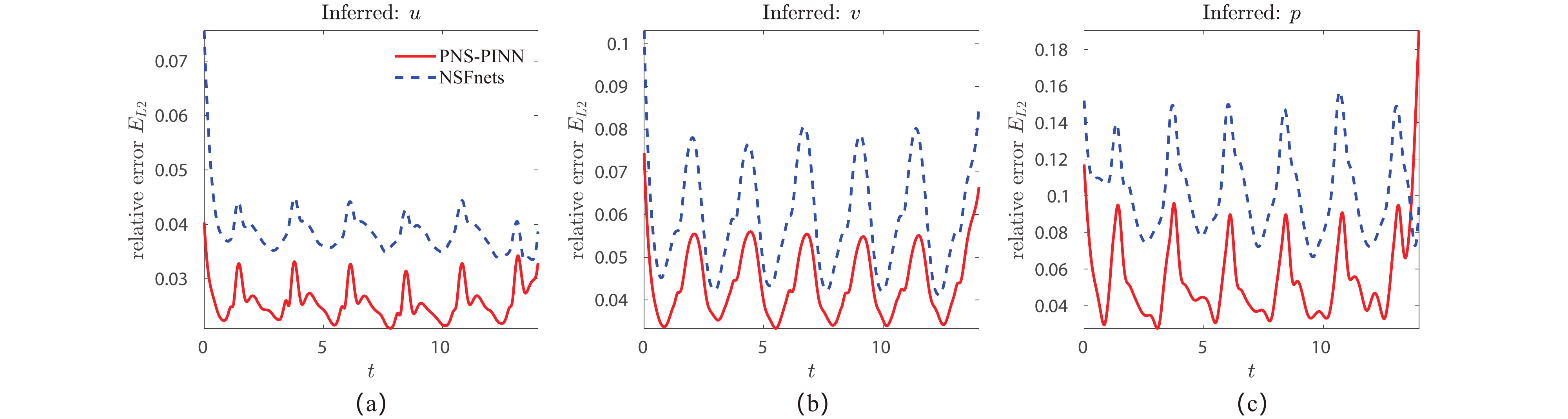}
  \caption{Relative error at convergence of the PNS-PINN and the NSFnets using velocity dataset with SST $k-\omega$ model: (a) velocity $u$; (b) velocity $v$; (c) pressure $p$.}
  \label{Fig:5}
\end{figure}

Fig.~\ref{Fig:5} illustrates the relative errors for PINN trained velocity components and pressure at convergence. It can be seen that the PNS-PINN is capable of inferring the flow with a relative error level of 2$\sim$5\% overally. The errors via the PNS-PINN are smaller and approximately in half of those obtained via the NSFnets. 
\begin{figure}[!htb]
  \centering
    \includegraphics[trim= 25mm 10mm 30mm 10mm,clip, width=0.8\textwidth]{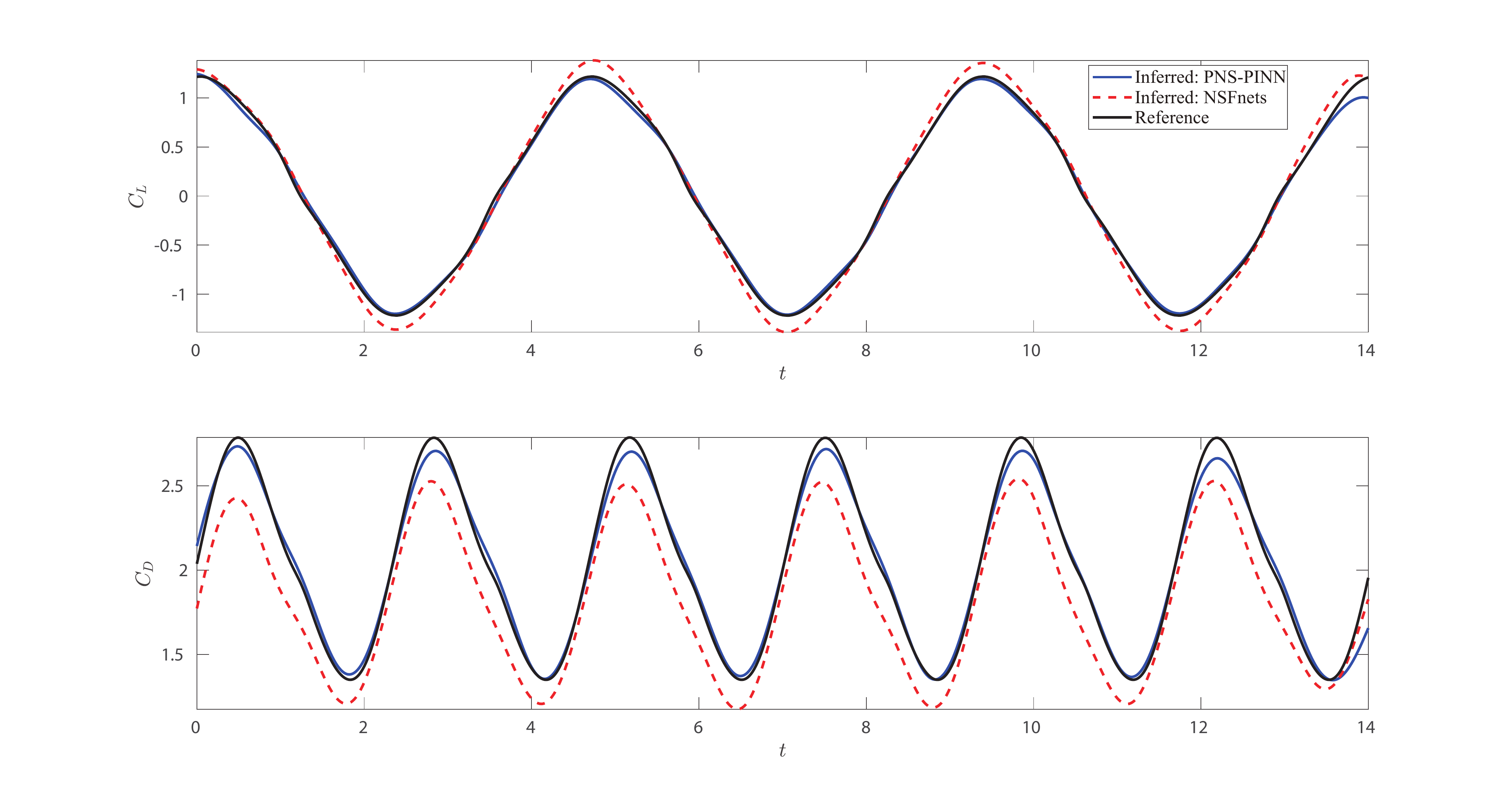}
  \caption{Inferred drag and lift coefficients via the PNS-PINN and the NSFnets using the velocity dataset with SST $k-\omega$ model.}
  \label{Fig:6}
\end{figure}

Based on the inferred flow fields, the lift and drag force coefficients of the vibrating cylinder can be determined through integration of pressure and velocity gradient distributions over the cylinder wall, with the following expressions,
\begin{equation}
\begin{aligned}
     F_L =\oint_{\bm \Gamma} [ \frac{1}{Re}(\frac{\partial u}{\partial y} +\frac{\partial v}{\partial x}) n_x+\frac{2}{Re} \frac{\partial v}{\partial y} n_y -p n_y] \text{d}s,  \\
     F_D=\oint_{\bm \Gamma} [ \frac{2}{Re} \frac{\partial u}{\partial x} n_x+  \frac{1}{Re}(\frac{\partial u}{\partial y}+\frac{\partial v}{\partial x}) n_y-p n_x] \text{d}s,
\end{aligned}
  \label{Eq:7}
\end{equation}
where, $\bm \Gamma$ is the boundary of the cylinder wall; $n_x$ and $n_y$ are the $x$- and $y$-components of the unit normal vectors of the wall surface. The non-dimensional lift and drag coefficients are thus calculated with $\displaystyle C_{L,D}=F_{L,D} /(0.5\rho U_\infty^2 D)$, and $\rho$ is of unit value.

The predicted drag and lift coefficients obtained from the inferred flow field are shown in Fig.~\ref{Fig:6}. It shows that the peaks and troughs of the lift force coefficients are exaggerated via the NSFnets, while the PNS-PINN attains better agreement with the reference CFD result. Also, the drag force is much better inferred via the PNS-PINN. The large deviation of the predicted forces via NSFnets can be well interpreted if we refer to the unfavourable flow inference shown in Fig.~\ref{Fig:4}, especially the distorted pressure field in the near vicinity of the cylinder wall, as presented in above content. The deviation of the mean drag force from the reference data reflects an unacceptable background pressure field. The parameterised Navier-Stokes method, on the contrast, can well resolve the turbulent flow distribution, especially the pressure field, so that the force coefficients acting on the VIV cylinder can be much better predicted.

\begin{figure}[!htb]
  \centering
    \includegraphics[trim= 40mm 5mm 5mm 0mm,clip, width=1.0\textwidth]{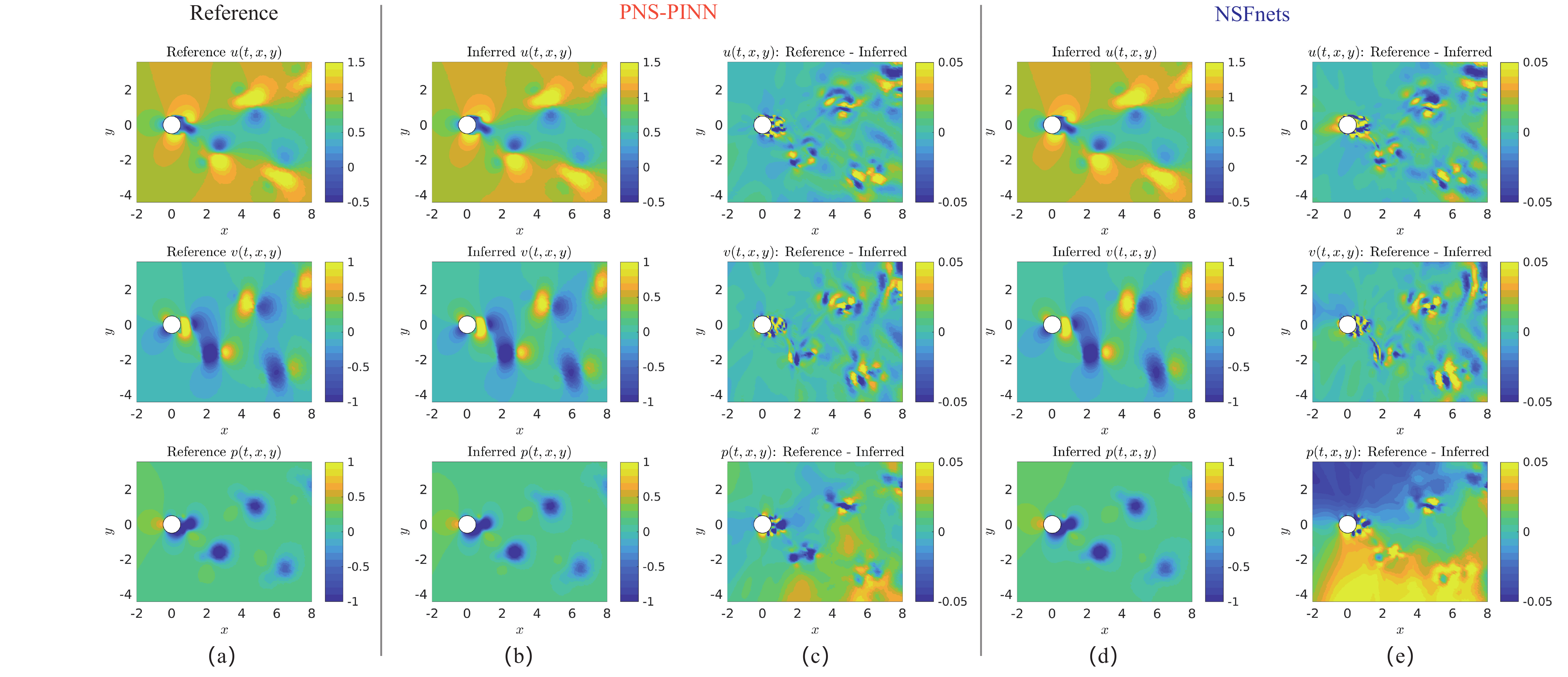}
   \caption{Flow fields inferred via the PINN and the velocity dataset with SST-SAS model: (a) reference data by CFD simulation; (b-c) inferred results via the PNS-PINN and the error distribution; (d-e) inferred results via the NSFnets and the error distribution.}
  \label{Fig:7}
  \end{figure}

Next, we move on to the machine learning of VIV in turbulent flow using the dataset with SST-SAS mode. Comparing with the  SST $k-\omega$ model, the SST-SAS  model performs better when dealing with flows with multiple scales, compared with the SST $k-\omega$ model. So, it is of further interesting to verify the effectiveness of reconstructing more complex flow characteristics via the PNS-PINN we applied in this work. The reference flow data provided by the CFD simulation at time instant $t$ = $\Delta t$/2 are shown in Fig.~\ref{Fig:7}(a). Together, Fig.~\ref{Fig:7}(b-c) and \ref{Fig:7}(d-e) show corresponding inferred flow distributions at $t$ = $\Delta t$/2. Via the PNS-PINN, visualization based on the inferred  velocity fields are quite satisfying, except that there shows comparatively obvious error at the head parts of the shedding vortices. As for the pressure distribution, larger error is visible not only at the head parts of the shedding vortices but also in the downstream region, especially at lower-right zone around $x \sim$ [4, 8]. One may also notice the errors of the inferred results for the dataset with SST-SAS model are overally larger than those those shown in Fig.~\ref{Fig:4} using the dataset with SST $k-\omega$. 

Additionally, the inferred flow distributions via the NSFnets show a larger discrepancy with the reference ones than the PNS-PINN, as is also found for the SST $k-\omega$ dataset. The errors of the velocity components, $u$ and $v$, are not only augmented at the vortex head parts but more importantly with much lower accuracy around the cylinder. As for the pressure distribution, the flow is poorly reconstructed especially in the near vicinity of the moving cylinder. 

\begin{figure}[!htb]
  \centering
    \includegraphics[trim= 30mm 0mm 35mm 0mm,clip, width=0.8\textwidth]{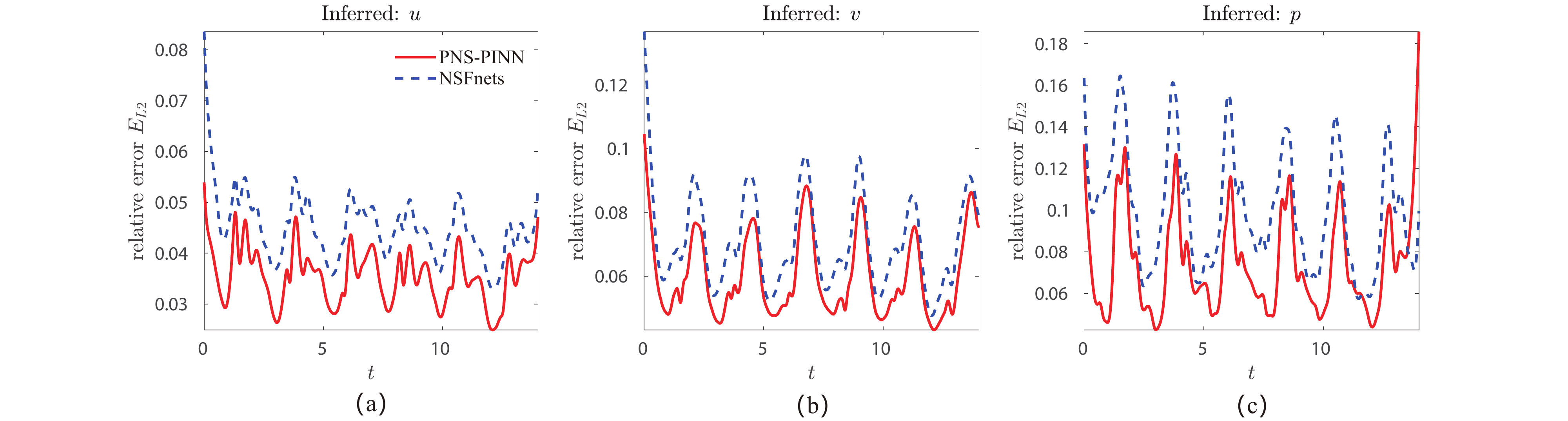}
  \caption{Relative error at convergence of the PNS-PINN and the NSFnets using velocity dataset with SST-SAS model: (a) velocity $u$; (b) velocity $v$; (c) pressure $p$.}
  \label{Fig:8}
\end{figure}

A more straightforward comparison of effectiveness between the two PINNs in this study for trainings using dataset with SST-SAS model can be found in Fig.~\ref{Fig:8}, where the the relative errors at convergence are illustrated. The errors for velocity distributions are around the order of 5-10$\times 10^{-2}$, and 15$\times 10^{-2}$ for the pressure field when the PNS-PINN are adopted. The error levels are much higher than PINN training results using dataset with SST $k-\omega$ model, as illustrated in Fig.~\ref{Fig:5}. On the other hand, the NSFnets provides flow results with larger convergence errors.

\begin{figure}[!htb]
  \centering
    \includegraphics[trim= 25mm 10mm 30mm 10mm,clip, width=0.8\textwidth]{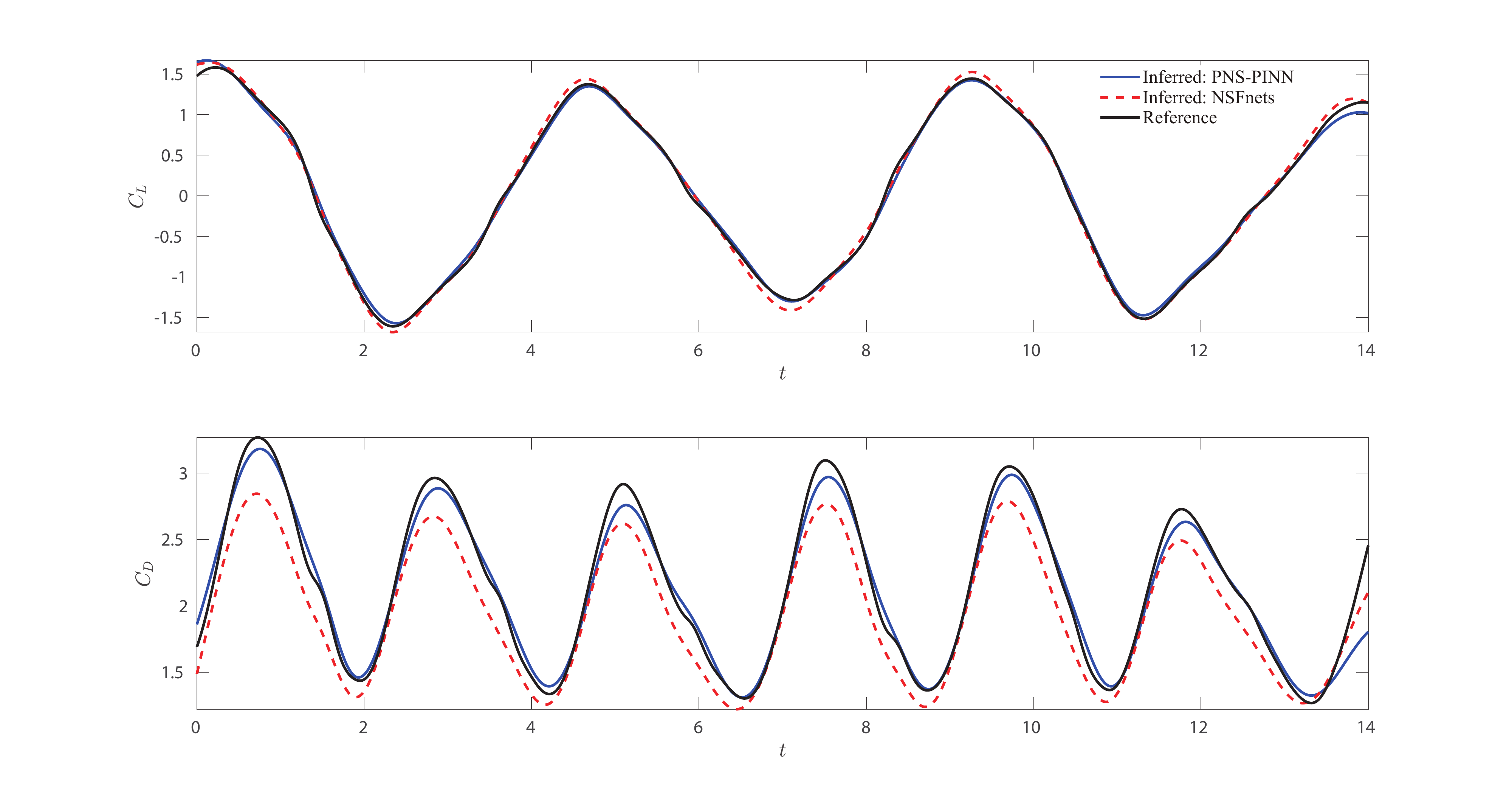}
  \caption{Inferred drag and lift coefficients via the PNS-PINN and the NSFnets using velocity dataset with SST-SAS model.}
  \label{Fig:9}
\end{figure}

The inferred force coefficients on the VIV cylinder were obtained and shown in Fig.~\ref{Fig:9}. The multi-scale motion of the flow obtained by the SST-SAS model CFD simulation induced multi-component variation of the drag and lift in time dimension, which is not obvious when we look into those by the SST $k-\omega$ simulation dataset in Fig.~\ref{Fig:6}. Again, the peaks and troughs of the lift force curve are exaggerated via the NSFnets, while the PNS-PINN delivers lift force with quite high accuracy throughout the time coverage. Via PNS-PINN, the results present improvement over the NSFnets configuration.

\subsection{Machine learning using dye trace dataset}
In this section, scattered dye trace data and the motion of the VIV cylinder are provided to the PINN, and the performance of the PNS-PINN and that of the NSFnets are evaluated. Here we assume that only the dye trace data behind the cylinder and the motion record are available. Our focus is to gain flow physics accompanying the VIV phenomenon with high fidelity in turbulent scenario. This is of importance, for example, for flow diagnosis either in laboratory tests or even large-scale engineering or natural events.

\begin{figure}[!htb]
  \centering
    \includegraphics[trim= 30mm 0mm 30mm 0mm,clip, width=0.8\textwidth]{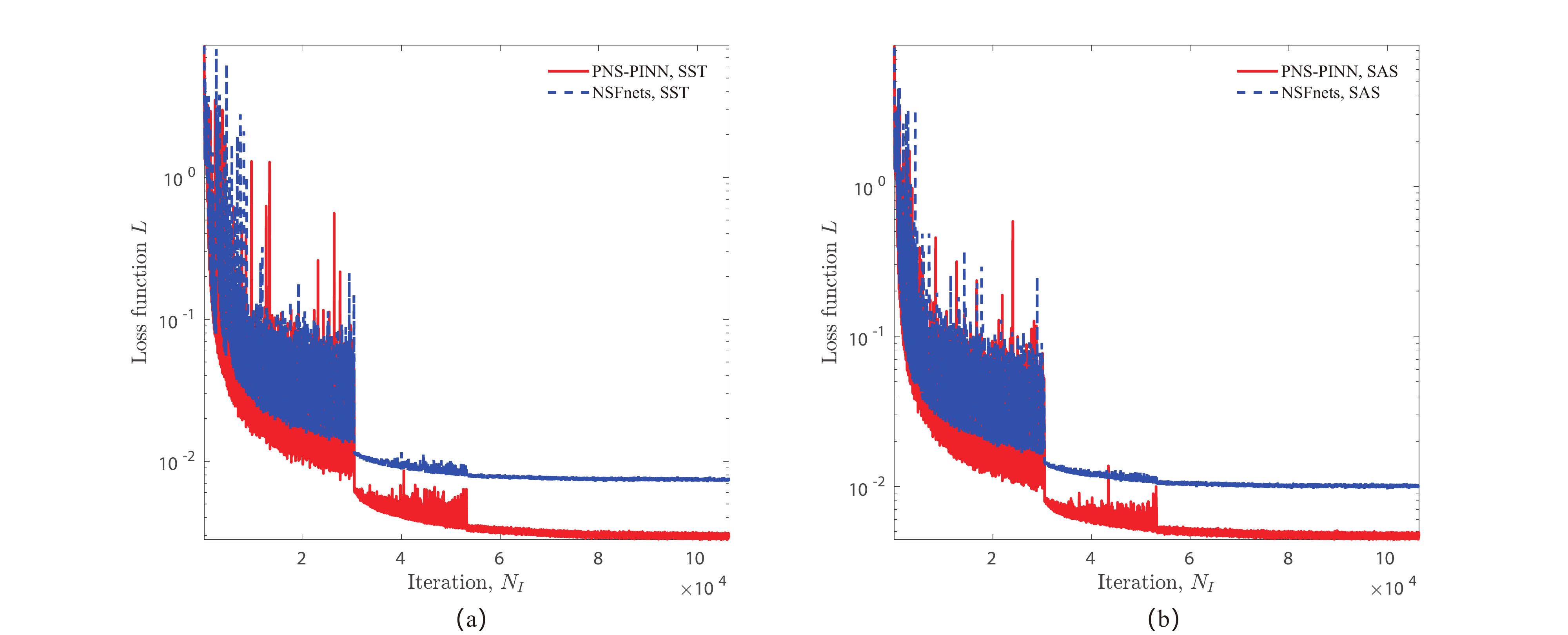}
   \caption{Time histories of the loss functions of (a) the PNS-PINN and (b) the NSFnets trainings using dye trace dataset.}
  \label{Fig:10}
\end{figure}
\begin{figure}[!htb]
  \centering
    \includegraphics[trim= 0mm 3mm 0mm 0mm,clip, width=1.0\textwidth]{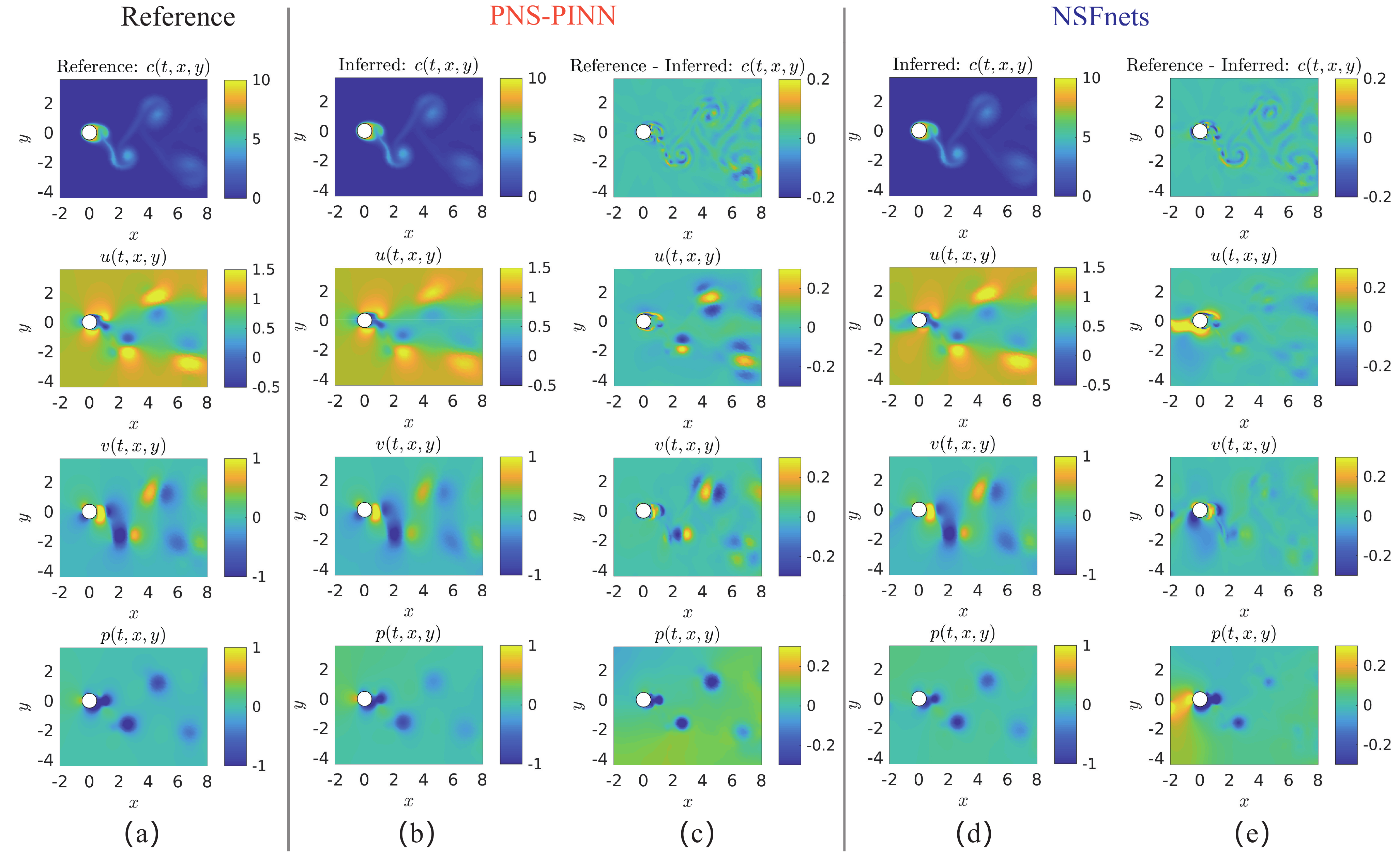}
   \caption{Flow fields inferred via the PINN and dye trace dataset with SST $k-\omega$ model: (a) reference data by CFD simulation; (b-c) inferred results via the PNS-PINN and the error distribution; (d-e) inferred results via the NSFnets and the error distribution.}
  \label{Fig:11}
\end{figure}   
  
Based on the SST and SST-SAS  CFD simulation dataset, the PINN training incorporating  the dye trace concentration $c$ and its governing transport equation are performed for both the parameterised and the normal NS configurations. The convergence history of the loss function is shown in Fig.~\ref{Fig:10}. The  initial loss values are larger than those in the velocity dataset based training shown in Fig.~\ref{Fig:3}, still they converge to steady levels as the learning rate gets smaller after $N_I> 3\times 10^4$. For the SST $k-\omega$ dataset, convergence loss values for the PNS-PINN and the NSFnets are 0.30\% and 0.74\%; Similar with the velocity data based training, loss function at convergence attain higher values using the SST-SAS dataset and are 0.47\% and 1.00\% respectively for the two PINNs.

\begin{figure}[!htb]
  \centering
    \includegraphics[trim= 30mm 0mm 35mm 0mm,clip, width=0.8\textwidth]{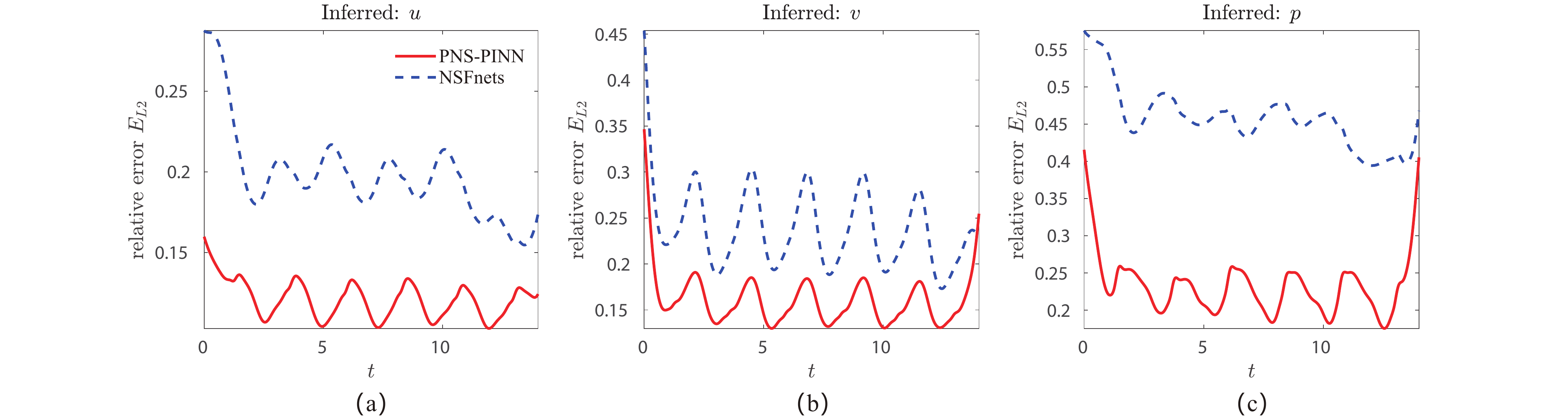}
  \caption{Relative error at convergence of the PNS-PINN and the NSFnets using dye trace dataset with SST $k-\omega$ model: (a) velocity $u$; (b) velocity $v$; (c) pressure $p$.}
  \label{Fig:12}
\end{figure}
\begin{figure}[!htb]
  \centering
    \includegraphics[trim= 25mm 10mm 30mm 10mm,clip, width=0.8\textwidth]{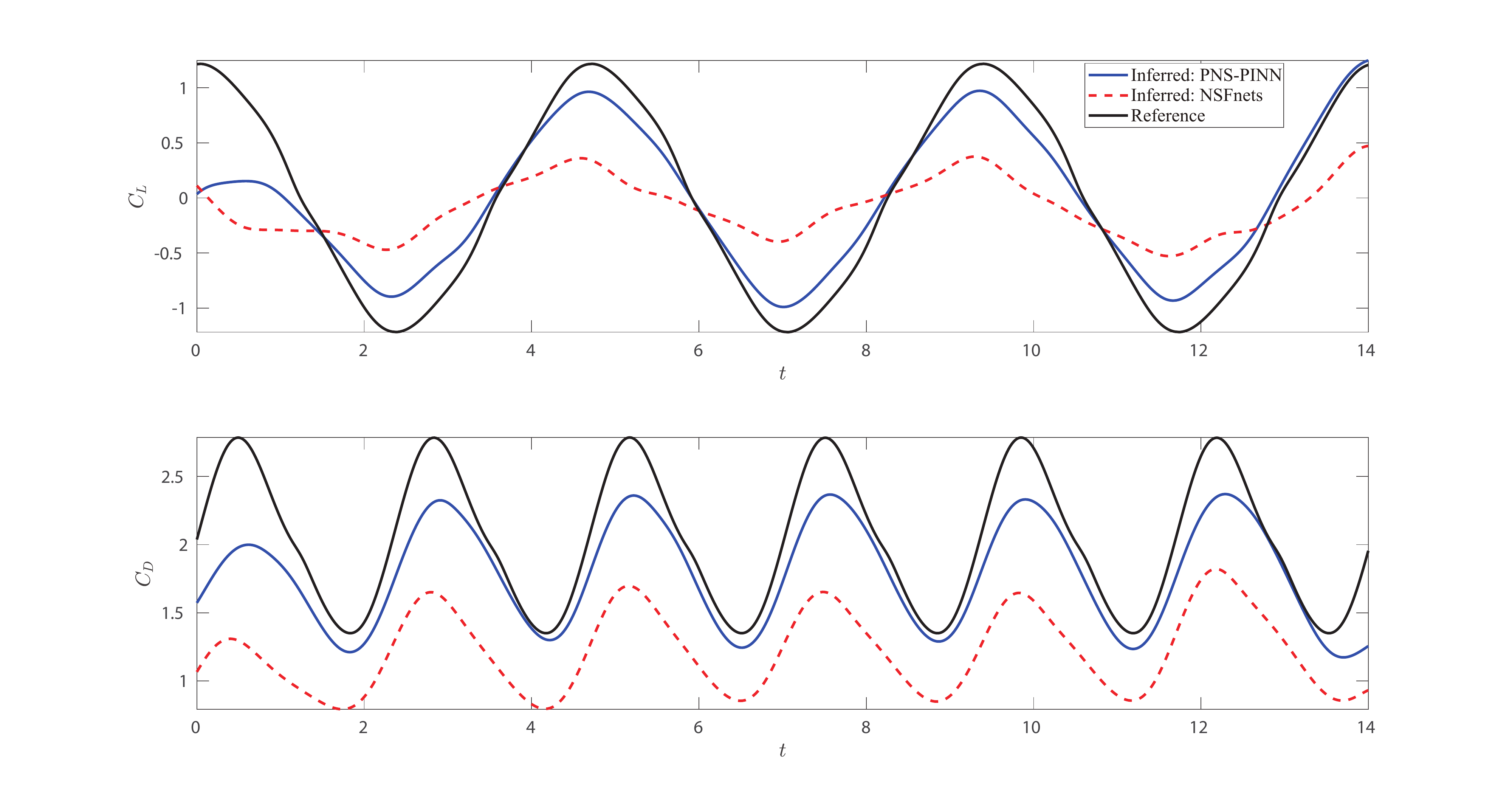}
  \caption{Inferred drag and lift coefficients via the PNS-PINN and the NSFnets using dye trace dataset with SST $k-\omega$ model.}
  \label{Fig:13}
\end{figure}

Fig.~\ref{Fig:11} illustrates the reference flow and the inferred results for both the PNS-PINN and NSFnets configurations, using the training dataset with SST $k-\omega$ model. Regarding the parameterised Navier-Stokes based PNS-PINN, the dye concentration together with the velocity and pressure fields can be well reproduced in aspect of visualization, as shown in Fig.~\ref{Fig:11}(b-c). Yet, compared with  those shown in Fig.~\ref{Fig:4} using the velocity dataset, one may notice that the errors of the velocity components are larger though with a similar error distribution pattern. The error for the pressure field is not only higher in magnitude, but also it has a background value that does not show up in Fig.~\ref{Fig:4}. It is speculated that the the discrepancy is mainly induced by the locally insufficient concentration data, for example in the upstream and the farfield regions of the cylinder.  The above comparison and findings demonstrate that, for VIV flow in turbulent regime, the flow inference obtained via the PNS-PINN using merely the dye trace can deliver satisfying visualisation, despite being less accurate to some extent than the training results directly using the velocity dataset. 
  
The inference of turbulent VIV flow via the NSFnets witnesses distortions in the near vicinity of the cylinder wall, both for the velocity components and the pressure, as pictured in Fig.~\ref{Fig:11}(d-e), although the inferred dye concentration is comparably as accurate as that by the PNS-PINN. The relative errors, illustrated in Fig.~\ref{Fig:12}, also shows that the pressure field inferred via NSFnets has been much distorted, having an error level over 40\%. The relative errors for $u$ and $v$ via the NSFnets are also higher than 20\%. On the contrast, the PNS-PINN delivers approximately half of error level for each flow field data via NSFnets.
\begin{figure}[!htb]
  \centering
    \includegraphics[trim= 0mm 3mm 5mm 0mm,clip, width=1.0\textwidth]{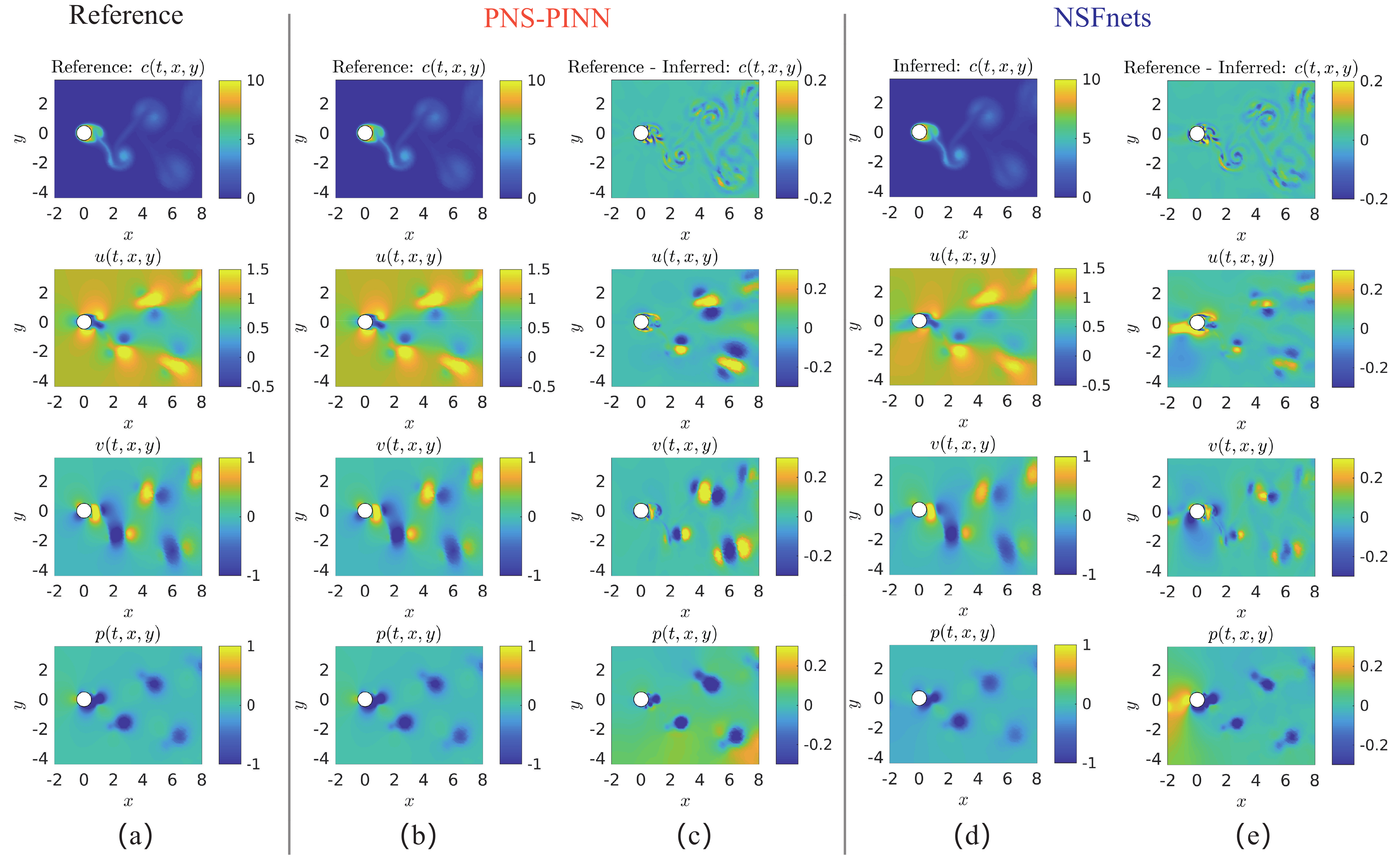}
   \caption{Flow fields inferred via the PINN and the dye trace dataset with SST-SAS model: (a) reference data by CFD simulation; (b-c) inferred results via the PNS-PINN and the error distribution; (d-e) inferred results via the NSFnets and the error distribution.}
  \label{Fig:14}
\end{figure} 

The accuracy of the inferred results analyzed above will consequently affect the prediction of forces acting on the VIV cylinder. Fig.~\ref{Fig:13}. The lift force via the PNS-PINN configuration generally holds up to the reference result except that the amplitude is underestimated by around 20\%, which is quite close to the loss level of the pressure field as shown in Fig.~\ref{Fig:12}. For the NSFnets, the amplitude and also the mean value of the drag force coeffcient via the PNS-PINN are both underestimated. The lower mean drag force indicates there exist a deviation of the mean pressure field (or most responsibly) between the inferred and the reference flows, as can be seen in Fig.~\ref{Fig:11}. The lower mean drag result by the NSFnets can be readily understood when we refer to the inferred pressure as well.
 
\begin{figure}[!htb]
  \centering
    \includegraphics[trim= 30mm 0mm 35mm 0mm,clip, width=0.8\textwidth]{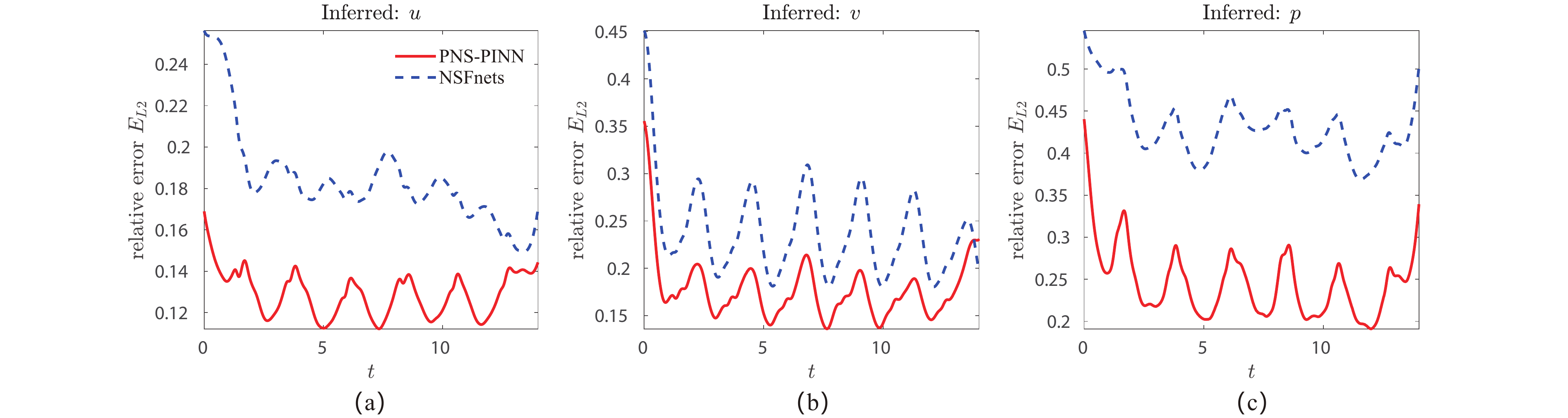}
  \caption{Relative error at convergence of the PNS-PINN and the NSFnets using dye trace dataset with SST-SAS model: (a) velocity $u$; (b) velocity $v$; (c) pressure $p$.}
  \label{Fig:15}
\end{figure}

\begin{figure}[!htb]
  \centering
    \includegraphics[trim= 25mm 10mm 30mm 10mm,clip, width=0.8\textwidth]{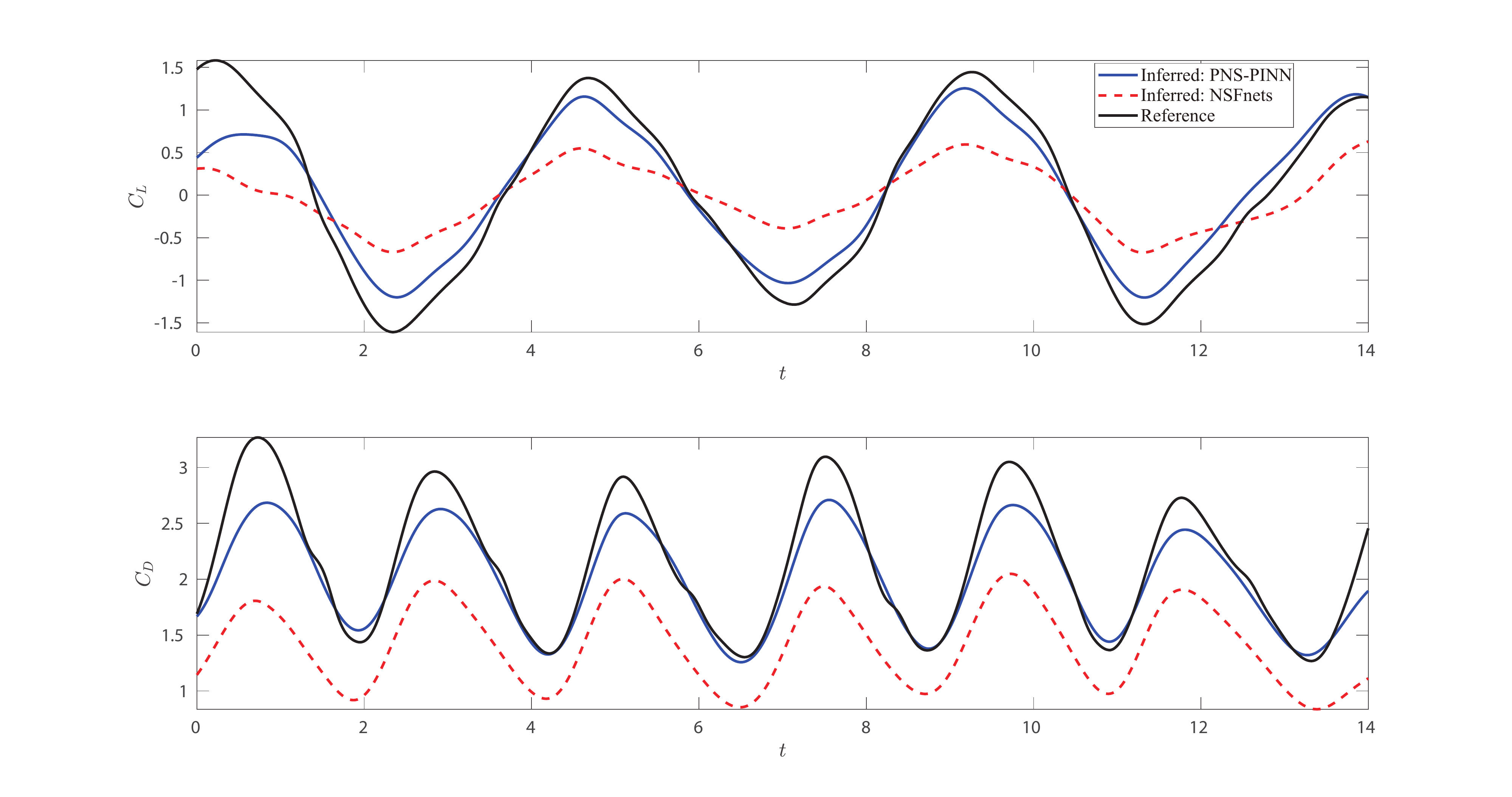}
  \caption{Inferred drag and lift coefficients via the PNS-PINN and the NS-PINN using dye trace dataset with SST-SAS model.}
  \label{Fig:16}
\end{figure}

The inferred flow data based on the dye trace dataset using the SST-SAS mixed model simulation have been also obtained. The results concerning the flow inference, reconstruction, error evaluation and the force coefficients on the VIV cylinder itself have been illustrated in Figs.~\ref{Fig:14}~-~\ref{Fig:16}. More detailed comments are omitted here for simplicity, as the main findings and conclusions for the training using dataset with SST-SAS model are in close accordance with those using dataset with SST model. Minor error level differences can be observed between the results by the two datasets.
 
\subsection{Discussion}

In the above section, totally eight cases, with different PINN configurations and dataset sources, were performed. Results for flow data inference and forces on VIV cylinders were reported. To further evaluate the effectiveness of the PINNs for VIV in turbulent flows, a summary of these cases and comparison between them, are made here.

The loss convergence $L$ and relative error levels $E_{X}$ are listed in Table.~\ref{Tab:2}. The relative error function $E[.]$ is defined in Eq.~(\ref{Eq:6}), while $E_{X}$ in the table are averaged results of the relative error over a time range of $t \sim $ [2, 12]. 

The PNS-PINN training results are presented in bold font. Taking trainings using dataset with the SST $k-\omega$ model for example, based on velocity, the flow data inferred via the PNS-PINN has the best convergence performance and the smallest relative errors, overally below 5\%.  The forces can also be accurately inferred. Based on the dye trace data, the loss convergence of the PNS-PINN training can be delivered with comparable level, despite that the relative errors of flow distributiobs are larger than those obtained using directly the velocity data. Specially the error in pressure reconstruction (up to 22.32\%) induces consequently considerable errors of the forces acting on the VIV cylinder. The NSFnets training generally doubled the loss and the error level of the inferred flow data based on velocity, while it brought 4-5 times larger error concerning the forces on cylinders undergoing VIV. This demonstrates successful application of the parameterised Navier-Stokes equation based PINN for VIV studies in turbulent flow. Flow inference via the PNS-PINN based on the dye trace, though not accurate as based on velocity data, can also provide a rough estimate of the flow and the forces in cases where the velocity data are not available. 

The training results by dataset with SST-SAS model indicate that, for turbulent flow with more unsteady characteristics, the PNS-PINN configuration in this study still proves to be qualified in VIV studies, though slightly higher error levels are found compared with that using dataset with SST $k-\omega$ model.
\begin{table}[!htb]
  \centering
  \caption{A summary on loss level and relative errors of inferred quantities via the PNS-PINN and the NSFnets trainings for VIV turbulent flow fields and force coefficients of the cylinder.}
    \begin{tabular}{ccccccccc}
    \toprule
    \multicolumn{9}{c}{Dataset: SST $k$-$\omega$} \\
    \midrule
    Dataset & PINN type & $L$ & $E_u$ & $E_v$ & $E_p$ & $E_c$ & $E_{C_D}$ &  $E_{C_L}$ \\
    \midrule
    \multicolumn{1}{c}{\multirow{2}[2]{*}{velocity}} & \textbf{PNS-PINN} & \textbf{0.21\%} & \textbf{2.52\%} & \textbf{4.39\%} & \textbf{4.97\%} & \textbf{--} & \textbf{2.03\%} & \textbf{3.45\%} \\
          & {NSFnets} & {0.62\%} & {3.85\%} & {5.93\%} & {10.05\%} & {--} & {10.20\%} & {11.87\%} \\
    \hline
    \multicolumn{1}{c}{\multirow{2}[2]{*}{dye trace}} & \textbf{PNS-PINN} & \textbf{0.30\%} & \textbf{11.81\%} & \textbf{15.66\%} & \textbf{22.32\%} & \textbf{1.39\%} & \textbf{12.55\%} & \textbf{24.54\%} \\
          & {NSFnets} & {0.74\%} & {19.50\%} & {23.86\%} & {45.40\%} & {1.94\%} & {39.30\%} & {72.90\%} \\
    \midrule
    \multicolumn{9}{c}{Dataset: SST-SAS} \\
    \midrule
    Dataset & PINN type & $L$ & $E_u$ & $E_v$ & $E_p$ & $E_c$ & $E_{C_D}$ &  $E_{C_L}$ \\
    \midrule
    \multicolumn{1}{c}{\multirow{2}[2]{*}{velocity}} & \textbf{PNS-PINN} & \textbf{0.38\%} & \textbf{3.49\%} & \textbf{5.98\%} & \textbf{7.04\%} & \textbf{--} & \textbf{3.18\%} & \textbf{3.70\%} \\
          & {NSFnets} & {0.78\%} & {4.42\%} & {7.01\%} & {9.58\%} & {--} & {9.53\%} & {5.54\%} \\
    \hline
    \multicolumn{1}{c}{\multirow{2}[2]{*}{dye trace}} & \textbf{PNS-PINN} & \textbf{0.47\%} & \textbf{12.49\%} & \textbf{16.98\%} & \textbf{23.30\%} & \textbf{1.91\%} & \textbf{8.54\%} & \textbf{21.19\%} \\
          & {NSFnets} & {1.00\%} & {18.06\%} & {23.46\%} & {41.95\%} & {2.43\%} & {32.43\%} & {62.28\%} \\
    \bottomrule
    \end{tabular}%
  \label{Tab:2}%
\end{table}%
\begin{figure}[htb] 
	\centering 
	\includegraphics[trim= 0mm 10mm 0mm 0mm,clip, width=\linewidth]{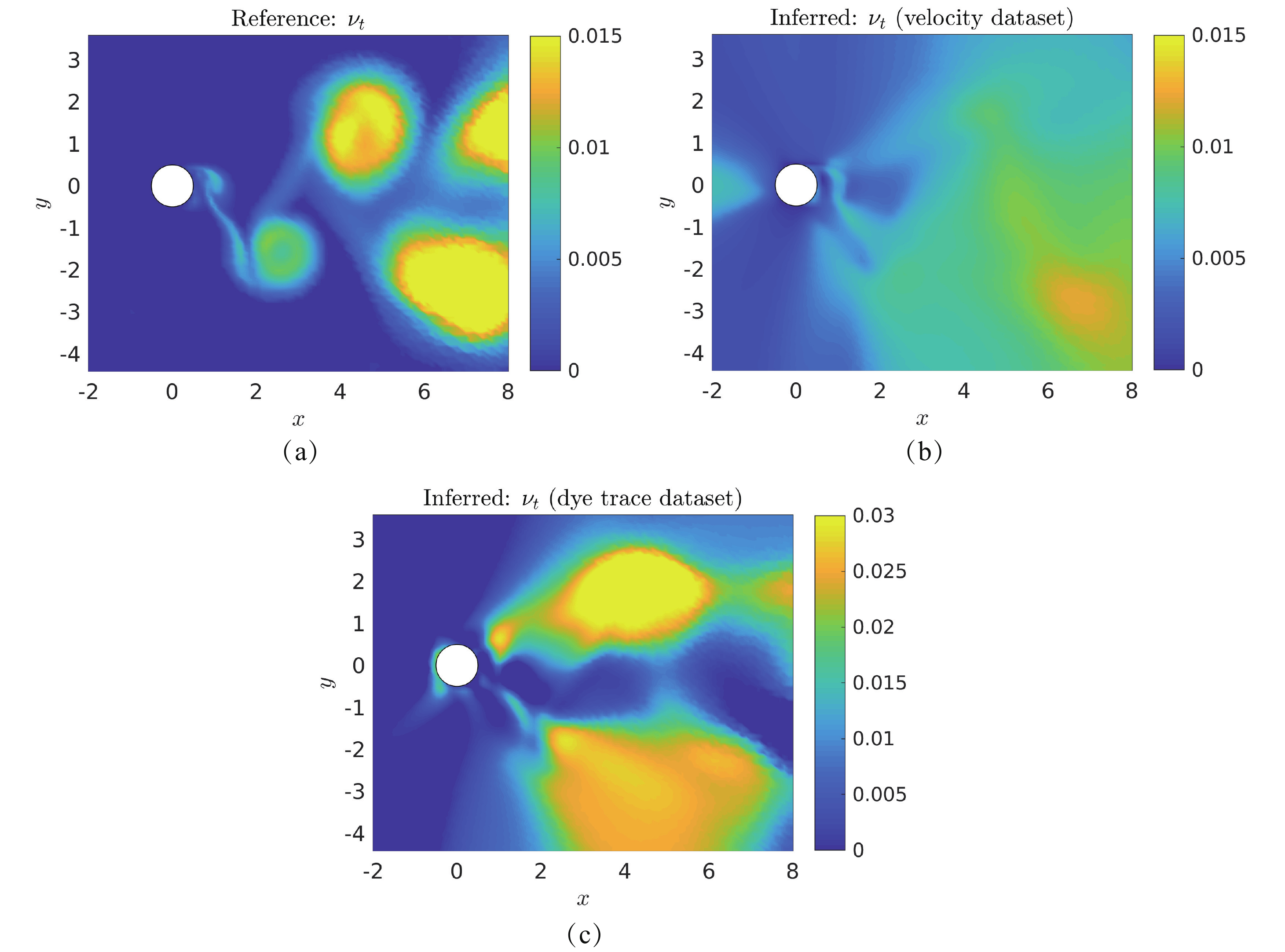}
      \caption{Inferred $\nu_t$ by the PNS-PINN using dataset with SST $k-\omega$ model: (a) reference distribution; (b) inferred distribution using the velocity dataset; (c) inferred distribution using the dye trace dataset.}
        \label{Fig:17}
\end{figure} 

In the PNS-PINN configuration we applied in this study for turbulent flows passing cylinder undergoing VIV, the hidden variable $\nu_t$ is inferred through the machine learning using velocity or dye trace dataset, cylinder motion and corresponding physics constraints, \textit{i.e.}, governing partial differential equations. The analysis in preceding sections shows that, based on limited and scattered velocity or dye trace and cylinder motion data, the parameterised Navier-Stokes equation based physics informed neural network can be used to predict the flow distribution, especially the pressure data that is not so straightforward to obtain, as well as the forces acting on cylinder undergoing VIV motions in turbulent flows, \textit{e.g.}, in experimental occasions. 

Here, we will explore the artificial viscosity that we introduced in the parameterised Navier-Stokes equation and its inferred distribution obtained by the PINN trainings. Fig.~\ref{Fig:17} (a) presented the reference eddy viscosity $\nu_t$ through the SST $k-\omega$ turbulence model CFD simulations. In Fig.~\ref{Fig:17} (b-c), the inferred values through the PNS-PINN training using the velocity and the dye trace datasets are pictured. 

One can tell that the inferred artificial viscosity $\nu_t$ does not match well with the reference distribution resolved by the CFD simulation. However, the distribution patterns are similar to some extent between the reference and the inferred artificial viscosity results. It is noticed that the low-value and the high-value $\nu_t$ regions match well with the reference one for results using both the velocity and the dye trace datasets. Also, in the regions where vortices take dominance, the PNS-PINN training underestimated the viscosity using the velocity dataset, while overestimated the viscosity using the dye trace dataset. Thus, the inference through the concentration transport introduces more viscosity effect. On the other hand, it is of interest when we consider the fact that conventional turbulence models seek calibrations by tuning model constants according to specified engineering practices. However, in the PNS-PINN applied in this study, the viscosity $\nu_t$ is introduced as a hidden output variable, and the determination of $\nu_t$ via PNS-PINN is free of special calibrations and model tunings. It indicates that, by implanting additional eddy viscosity, the PNS-PINN method can be applied in more universal and complex studies, compared with conventional turbulence models.

\section{Conclusions}
In this study, a parameterised Navier-Stokes equation based PINN, termed PNS-PINN, was employed to investigate the turbulent flow around a cylinder undergoing VIV motion with Reynolds number $Re$ = $10^4$. A Navier-Stokes equation based PINN, termed NSFnets, was also considered for comparison. The training datasets were prepared by the CFD simulations with the SST  and the SST-SAS turbulence models. Input of the both the PNS-PINN and the NSFnets includes scattered velocity and cylinder motion data, or dye trace and cylinder motion data. Results on the training convergences, the inferred flows and forces, were presented and discussed. Main conclusions of this study include:
\begin{itemize}

\item[(1)] The PNS-PINN, which introduces an artificial viscosity parameter $\nu_t$ as a hidden variable in the neural network and the physics constraint, performs better in loss convergence, flow inference and force prediction concerning the VIV in turbulent flow scenario. Specially, the pressure field is delivered with satisfying accuracy without any training data source itself.  The NSFnets, on the other hand, encounters lower accuracy in velocity inference and distortion in pressure field reconstruction. Consequently, the relative errors of the predicted forces via the PNS-PINN are around one third of those via the NSFnets.

\item[(2)] For the PNS-PINN, based on the velocity dataset, the relative errors of quantities of interest (\textit{i.e.}, velocity components $u$ and $v$, pressure $p$, force coefficients $C_D$ and $C_L$), are all below 5\% for turbulent VIV flow. The relative errors based on dye trace dataset are generally 10\%-17\% for the velocity components, 22\%-24\% for the pressure field, and 10\%-25\% for the force coefficients.

\item[(3)] The PNS-PINN can deal with flow inference for dataset with both the SST and the SST-SAS models, though slight larger error levels are found for the latter. This implies promising effectiveness of the PNS-PINN to deal with turbulent VIV flows.

\item[(4)]Unlike conventional turbulence modeling in CFD simulations, by implanting additional eddy viscosity, the PNS-PINN employed in this study is free of extra calibaration and tunings. In future, more applications of the PNS-PINN for flow data assimilationc are expected, such as in circumanstances of experimental measurement and flow diagnosis, \textit{etc.}
\end{itemize}

\section{Acknowledgements}
The authors owe great appreciation to Dr. Yong Wang in Max Planck Institute for Dynamics and Self-Organization, for his constructive advice and all the effort in manuscript preparation. This work was financially supported by the National Natural Science Foundation of China (Grant Nos. 51809084,  91851127), the National Science Fund for Distinguished Young Scholars of China (Grant No. 51425901), the Fundamental Research Funds for the Central Universities of China (Grant No. B200202049).
\bibliographystyle{unsrt} 
\bibliography{references}

\end{document}